\documentclass[pre,twocolumn,showpacs,aps,floatfix,superscriptaddress]{revtex4}

\usepackage[pdftex]{graphicx}
\usepackage{multirow}
\usepackage{color}
\usepackage{amssymb,amsfonts,amsmath}
\usepackage[english]{babel}

\begin{document}

\title{Information filtering in complex weighted networks}

\author{Filippo Radicchi}\affiliation{Amaral Lab, Chemical and Biological Engineering,
Northwestern University, Evanston, IL, USA}

\author{Jos\'e J. Ramasco}
\affiliation{Instituto de F\'{\i}sica Interdisciplinar y Sistemas
  Complejos IFISC (CSIC-UIB), E-07122 Palma de Mallorca, Spain}
\affiliation{Complex Networks \& Systems Lagrange Laboratory, ISI Foundation, Turin, Italy}

\author{Santo Fortunato}\affiliation{Complex Networks \& Systems Lagrange Laboratory, ISI Foundation, Turin, Italy}

\begin{abstract}
Many systems in nature, society and technology can be described as
networks, where the vertices are the system's elements and edges
between vertices indicate the interactions between the corresponding
elements. Edges may be weighted if the interaction strength is
measurable. However, the full network information is often
redundant because tools and techniques from network analysis do
not work or become very inefficient if the network is too
dense and some weights may just reflect measurement errors, and
shall be discarded. Moreover, since weight distributions in many
complex weighted networks are broad, most of the weight is
concentrated among a small fraction of all edges. It is then crucial to properly detect relevant edges.
Simple thresholding would leave only
the largest weights, disrupting the multiscale structure of the system,
which is at the basis of the structure of complex networks, and ought
to be kept. In this paper we propose a weight filtering
technique based on a global null model (GloSS filter), keeping both the weight distribution and the full topological
structure of the network. 
The method correctly quantifies the statistical significance
of weights assigned independently to the edges from a given distribution.
Applications to real
networks reveal that the GloSS filter is indeed able to identify relevant
connections between vertices.
\end{abstract}


\maketitle

\section{Introduction}
A popular way to look at a complex system is turning it into a
{\it graph}, or {\it network}, by highlighting the fundamental
elements of the system (vertices) and the interactions between them
(edges connecting vertices), possibly with their
strength (weights on edges). Due to the recent availability of massive data
sets and computational facilities capable to process them, many
networked systems have been carefully investigated 
in the last years~\cite{albert02,boccaletti06,newman03,dorogovtsev01,pastor04,barrat08,caldarelli07}. 

A recurrent property is the heterogeneity in
the distributions of the main structural features of such
systems. These include purely topological attributes,
like the number of neighbors of a vertex (degree)~\cite{albert99,Barabasi:1999} as well as
variables depending on the weighted character of the edges, like the
edge weights and the sum of the weights of the edges incident on a
vertex (strength)~\cite{Barrat:2004b}. Such heterogeneity is
responsible for peculiar properties of complex networks, like their
high robustness against random attacks or failures~\cite{cohen00}.
Weights and topology are by no means independent, revealing a
set of non-trivial relationships~\cite{Barrat:2004b}.
For this reason it is improper to separate weights from topology and to study the
system by exploiting either source of information.

However, keeping the full information about the network can give rise
to problems. A large network with a high edge density may be
intractable by traditional tools of network analysis. For instance, it
may be impossible to produce a meaningful visualization of the
network. Also, a high edge density is a serious obstacle for graph
clustering techniques~\cite{fortunato2010}, most of which rely on the
working assumption that the network is sparse, i.e. that the number of
edges is not much larger than the number of vertices.  Other analysis
tools may not be applicable due to their high computational
complexity. In addition, the estimates of the edge
weights may be biased by measurement errors, so the connections
between some pairs of vertices might not be meaningful.

For all these reasons, it is important to develop suitable techniques 
to reduce the network, by maintaining only the most valuable information. The problem of information
reduction in datasets has a long tradition and has led to the design
of very popular methods, like Principal Component Analysis~\cite{Jolliffe:2002}. 
For networked data a well known strategy is coarse graining~\cite{Kim:2004,Itzkovitz:2005,song05,Gfeller:2007},
which consists in grouping vertices based on their mutual similarity
or topological role in the network and replacing each group with
super-vertices. Here, instead, we wish to preserve all vertices and
act only on the edges, by selecting the most relevant ones.
This is a major challenge. For one thing, it should be clarified
what ``relevant'' means, as this is not straightforward. In
fact, several options are possible, depending on the features of the 
system that shall be preserved. Since edge weights are usually
broadly distributed, keeping just the largest
weights is a viable option, since a few edges account for most of the
total weight. 
All weights lower than a predefined
threshold could be then erased~\cite{farkas07,Eguiluz:2005,Allesina:2006,wu:2006,ramasco:2007}. 
However, global thresholding has two drawbacks. On the one hand, it introduces a scale
in an originally multi-scale system. On the other hand it may spoil
important topological properties. For instance, it may fragment the
network into a large collection of components. To avoid that, one may
construct a {\it maximum spanning tree}~\cite{Macdonald:2005}, where 
as many edges as possible are removed such to maintain the connectedness of the
graph and to keep the largest possible total weight on the remaining
edges. This traditional technique is also not ideal, as it reduces the
network to an acyclic graph (a tree), whereas cycles are very
important structural features of complex networks. Moreover, a tree
has a number of edges equal to the number of vertices minus one, and
it is unlikely that the number of relevant edges simply depends on the
number of vertices, for any system. Tumminello {\it et al}. have shown
that many more edges/information can be kept, by
extracting a subgraph that can be embedded on a surface of genus $k$,
instead of a tree~\cite{tumminello05}. 

Still, selecting edges with a systematic bias towards the largest
weights would destroy the heterogeneity in the distribution of edge
weights, which is a crucial feature of complex weighted
networks. Furthermore, this could significantly modify the 
coupling between weights and topology.  
Meanwhile there are a few methods capable
to filter the information on the edges such to
respect the multiscale structure of complex weighted
networks. Such techniques include a two-stage algorithm proposed by
Slater~\cite{slater09,slater10} and a method by Glattfelder and
Battiston~\cite{glattfelder09} based on a multilevel network analysis. 
In recent works by Serrano {\it et al}.~\cite{serrano09,serrano09a} the focus is on the immediate neighborhood of each
vertex.
For a given vertex, the weights on its
adjacent edges are analyzed, and those edges
carrying a significant fraction of the total strength of the
vertex are picked. The significance of the weight is estimated from the so-called
{\it disparity function}, that results from a
simple null model stating how weights are distributed among the
edges incident on the vertex. 
Here we focus on the edges, i.e. on pairs of connected vertices,
rather than on the individual vertices.
Unfortunately, it is not
possible to treat pairs of connected vertices independently of the
rest of the network, as they are attached to
other vertices, etc. 
The natural solution is a global null model, that
accounts for the full topology of the network, while preserving the
heterogeneity of the weight distribution. In this paper we propose the Global
Statistical Significance (GloSS) filter, which satisfies these constraints.

At variance with other techniques, the GloSS filter yields a well 
defined global $p$-value for all edge
weights of the network. Furthermore, it correctly identifies
situations in which all edges are equally relevant/irrelevant, like when
weights are independently and identically distributed on
the edges. Finally, the performance of the GloSS filter on several 
real networks,
both directed and undirected, is compared with that of 
other filtering techniques.

\section{Results and Discussion}

\subsection{The GloSS filter}

The starting point is the weight matrix ${\bf W}$, whose element
$w_{ij}$ indicates the weight of the edge joining vertices $i$ and
$j$. If there is no edge/interaction between $i$ and $j$,
$w_{ij}=0$. The number of neighbors of vertex $i$ is its degree $k_i$.
We also recall that the strength~\cite{Barrat:2004b} $s_i$ of vertex $i$ is 
the sum of the
weights of the edges incident on $i$: $s_i=\sum_{j}w_{ij}$.
Our null model is a graph where the
connections of the original network are locked, while weights
are assigned to the edges by randomly extracting
values from the observed weight distribution $P_{obs}\left(w\right)$.  
This null model thus preserves
both the topology and the weight distributions of the original
network, by construction.  

Suppose that we want to evaluate the 
statistical significance, according to this null model, of the edge between
vertices $i$ and $j$, with observed weight
$w_{ij}$. The degrees and strengths of $i$ and $j$ are $k_i$, $k_j$,
$s_i$ and $s_j$. 
This can be formalized by means of a Bayesian approach. The
probability to observe weight $w_{ij}\neq 0$ on the edge, given the degrees and
strengths of its endvertices, reads
\begin{equation}
P\left(w_{ij}\left| s_i, k_i, s_j, k_j \right.\right) = P_{obs}\left(w_{ij}\right)\,  \frac{P\left(s_i,s_j\left| w_{ij}, k_i, k_j \right.\right)}{P\left(s_i,s_j\left| k_i,k_j\right.\right)} .
\end{equation}
The denominator on the right hand side is a normalization factor, while
$P_{obs}\left(w_{ij}\right)$ is a well defined number. 
In order to estimate the term in the numerator we must take into
account that $w_{ij}$, $k_i$, $k_j$ 
are given and so the "free" variables contributing to $s_i$ and $s_j$ are the weights of the remaining 
$k_i-1$ and $k_j-1$ connections of vertices $i$ and $j$, respectively.
These weights can be treated as independent 
random variables
in the null model, with the only restrictions
that $\sum_{k \neq j} w_{ik} = s_i - w_{ij}$
and $\sum_{k \neq i} w_{jk} = s_j - w_{ij}$. This implies that
\begin{equation}
\begin{array}{ll}
P\left(s_i,s_j\left| w_{ij}, k_i, k_j \right.\right) =  &
F\left(s_i-w_{ij}, k_i-1\right) \times
\\
&  F\left(s_j-w_{ij}, k_j-1 \right)
\end{array} \; .
\label{eq:prob}
\end{equation}  
The function $F\left(s, k\right)$ is  the probability of randomly extracting, from the weight distribution
$P_{obs}\left(w\right)$, $k$ elements whose sum is equal to $s$, which means that
 \begin{equation}
\begin{array}{ll}
F\left(s, k\right) =  & \int dx_1 P_{obs}\left(x_1\right) \, \int P_{obs}\left(x_2\right) dx_2 \cdots
\\
& \cdots \int dx_k P_{obs}\left(x_k\right) \;
\delta\left(x_1+x_2+\ldots+x_k-s \right) 
\end{array} \; ,
\label{eq:func}
\end{equation}  
where the Dirac delta $\delta\left(x_1 +\ldots+x_k-s\right)$ ensures 
the satisfaction of the constraint on the vertices' strength.
We remark that, if either $i$ or $j$ (or both) has degree $1$,
Eq.~\ref{eq:prob}, as it stands, would not be defined. Here the whole
strength of $i$ (or $j$) would come from the edge $ij$, 
so the probability distribution of observing
that weight is just a $\delta$-function centered at $w_{ij}$, since no other values are
compatible with the strength of the vertex ($s_{i,j}=w_{ij}$ if
$k_{i,j}=1$).

Finally, the statistical significance (or $p$-value) $\alpha_{ij}$ 
of the observed
edge weight $w_{ij}$ can be computed by calculating
the integrals
 \begin{equation}
\begin{array}{ll}
\alpha_{ij} = P\left(>w_{ij}\left| s_i,k_i,s_j,k_j \right.\right) =
\frac{\int_{w_{ij}}^{\infty} dw \, P_{obs}(w)\, P\left(s_i,s_j\left| w, k_i, k_j \right.\right)}{\int_{0}^{\infty} dw \, P_{obs}(w)\, P\left(s_i,s_j\left| w, k_i, k_j \right.\right) }
\end{array}  \; .
\label{eq:pval}
\end{equation}  
Despite its apparently high complexity,
the computation of the significance level
can be carried out numerically in a fast
and accurate way. The probability function $F\left(s, k\right)$ can in fact be viewed
as a multiple convolution
integral of the weight distribution function and its computation 
may be performed by invoking the convolution theorem. First
the Fourier transform of the weight distribution is calculated, then its
$k$-th power; the final answer is obtained by computing the
Fourier antitransform of the result (see details in
Appendix A). The extension
of the former procedure to directed networks is straightforward. 
If $w_{ij}$ denotes the weight of the directed edge going from vertex $i$ to 
vertex $j$, it is sufficient to substitute in the former equations 
$k_i$ and $s_i$ with $k_i^{out}$ and $s_i^{out}$, respectively. For
vertex $j$ one ought to replace
$k_j$ with $k_j^{in}$ and $s_j$ with $s_j^{in}$. Once the
$p$-value of each edge has been determined, we can establish
a certain threshold and deem
the edges as significant if their $p$-values lie above that threshold. This procedure defines what we have called the GloSS filter.

\subsection{Tests on random weight distributions}

\begin{figure}
\begin{center}
\includegraphics[width=\columnwidth]{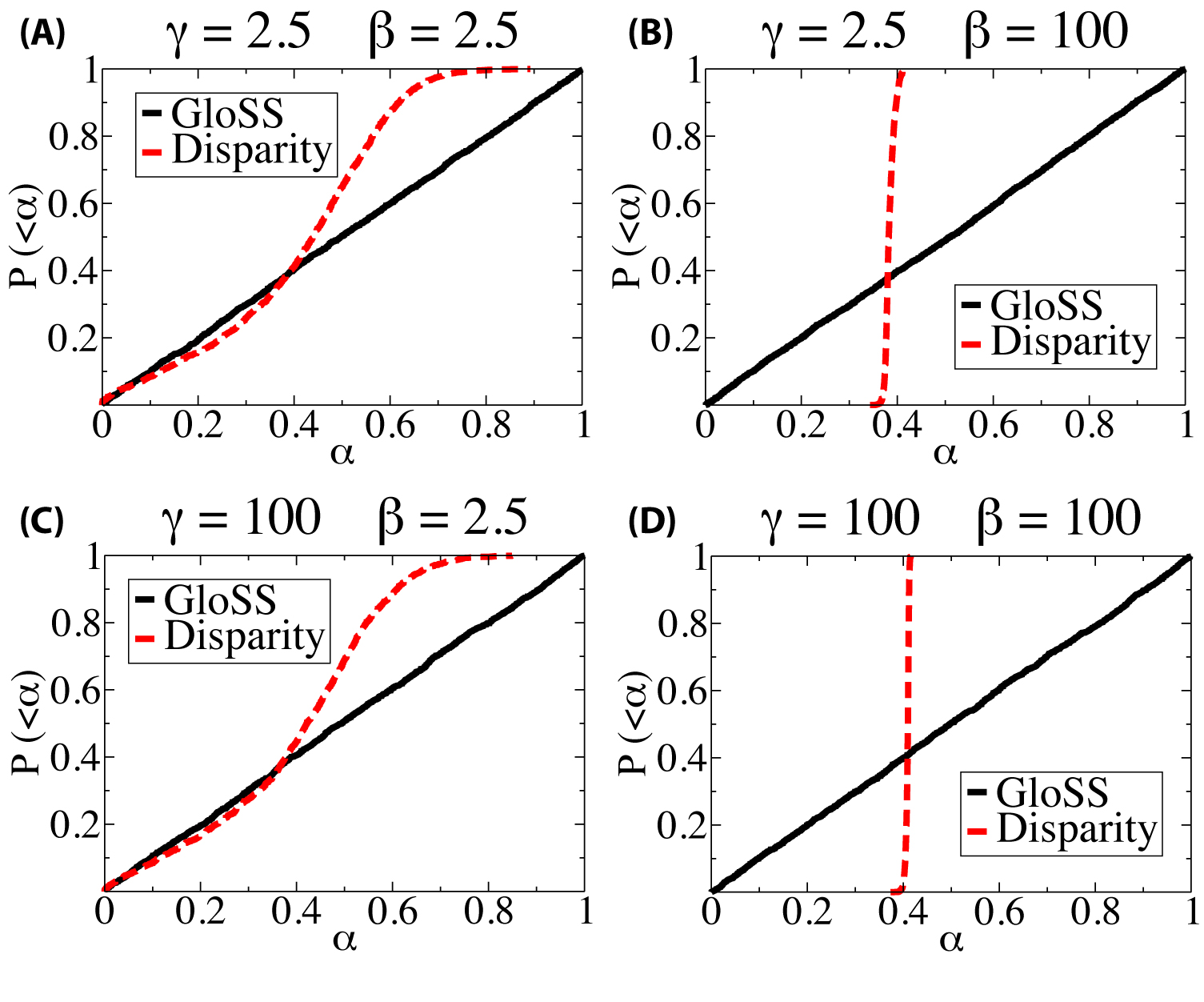}
\caption{\label{fig:null_model} (Color online) Cumulative distribution $P\left(<\alpha\right)$
of the significance level $\alpha$
for independent identically distributed weights.
Networks are made of
$N=1\,000$ vertices and have minimum degree 
equal to $5$. Connections among
vertices are randomly drawn by preserving
the {\it a priori} given degree sequence.
Vertex degrees and edge weights are
randomly chosen from the power law distributions
$P\left(k\right) \sim k^{-\gamma}$ and $P\left(w\right) \sim w^{-\beta}$,
respectively. Statistical significance of weights, for different choices of 
$\gamma$ and $\beta$, are computed with the GloSS filter
(continuous curve) and the Disparity filter
by Serrano {\it et al}.~\cite{serrano09} (dashed curve).}
\end{center}
\end{figure}

Ideally, any filtering procedure should be able to recognize
situations in which there are no significant weights. For instance,
given a distribution, we could assign weights taken from that
distribution on each edge, independently of the other edges. In this
way, the distribution of the weights on the edges would be random, with
no correlations with topological features. Therefore, 
the fluctuations of the weights coming from such distribution are
just the expected fluctuations of the distribution itself, whose
statistical significance is exactly indicated by the $p$-value $\alpha$
of Eq.~\eqref{eq:pval}. The probability $P\left(<\alpha\right)$ for an observed weight to have a
$p$-value $\alpha$ or lower is then exactly equal to $\alpha$, as all
$p$-values are equally probable. In
Fig.~\ref{fig:null_model} we show the profile of
$P\left(<\alpha\right)$ on random networks with power law distributions of
degrees and weights, with exponents $\gamma$ and $\beta$,
respectively. The four panels correspond to different choices of $\gamma$ and $\beta$.
For high values of the exponents (like $\gamma, \beta=100$) the power law distribution is
effectively exponential. In all cases we see that the GloSS filter
recovers the expected relation $P\left(<\alpha\right)=\alpha$
(diagonal continuous line), which indicates that indeed weights are
randomly distributed among the edges and there are no significant
fluctuations. The Disparity filter by Serrano {\it et al}.~\cite{serrano09},
instead, displays a different profile (dashed line). 
For actual power law
distributions of weights (Figs. 1A and 1C), it yields the expected pattern up to a
$p$-value of about $0.4$, then it deviates from it.
In particular, for the
case of exponential distributions of weights
(Figs.~\ref{fig:null_model}B and \ref{fig:null_model}D), all observed weights have
essentially the same $p$-value $\alpha \simeq 0.4$ (yielding the approximate step
function for the cumulative displayed in the figure). In this case the
values of the weights are quite close to each other, and the method
has problems to distinguish between them. We remark that, even if edge
weights are quite homogeneous here, once their distribution is defined
one can always assign to each weight a proper likelihood ($p$-value),
and discuss about its compatibility with the chosen distribution.
The different results obtained with the Disparity filter are due to
the different null model adopted by this filter, which is
local. However, at variance with the GloSS filter, it is not possible
to build a network based on the 
null model of the Disparity filter, just because of its local character. It is only possible
to restrict the picture to the subgraph consisting of a node and its
incident edges.

\begin{figure*}
\begin{center}
\includegraphics[width=\textwidth]{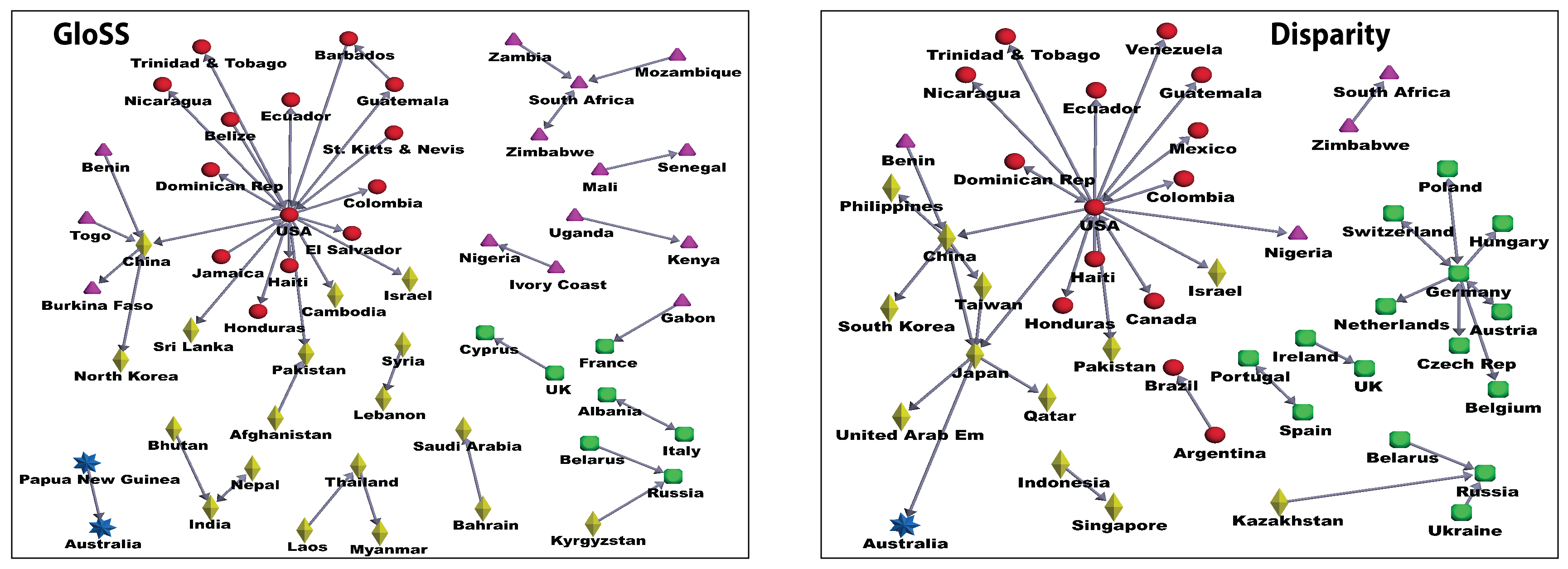}
\caption{\label{fig:trade} (Color online) Top $50$ connections of the World Trade
  Web in year $2006$: GloSS filter (left), Disparity filter
  (right). Countries without edges are removed from the picture.}
\end{center}
\end{figure*}
\begin{table*}
\begin{tabular}{r l  r r r r  c  r r r r}
& &  \multicolumn{4}{c}{GloSS} & & \multicolumn{4}{c}{Disparity}
\\
rank & & $\alpha$ & $w$ &  start vertex & end vertex & &  $\alpha$ &  $w$ & start vertex & end vertex
\\
\cline{3-6} \cline{8-11}
\\
$1$ & &$3 \times 10^{-7}$ & $4649.2$ & USA & Dominican Rep & & $4 \times 10^{-112}$ & $307823$ & USA & Canada 
 \\ 
 $2$ & &$1 \times 10^{-6}$ & $3893.1$ & USA & Honduras & & $1 \times 10^{-62}$ & $200515$ & USA & Mexico 
 \\ 
 $3$ & &$3 \times 10^{-6}$ & $520.17$ & Italy & Albania & & $3 \times 10^{-51}$ & $211247$ & Canada & USA 
 \\ 
 $4$ & &$5 \times 10^{-6}$ & $890.23$ & Haiti & USA & & $2 \times 10^{-34}$ & $4649.2$ & USA & Dominican Rep 
 \\ 
 $5$ & &$2 \times 10^{-5}$ & $1176.9$ & Zimbabwe & South Africa & & $4 \times 10^{-34}$ & $38386.1$ & USA & Venezuela 
 \\ 
 $6$ & &$3 \times 10^{-5}$ & $13084.3$ & Belarus & Russia & & $1 \times 10^{-33}$ & $13084.3$ & Belarus & Russia 
 \\ 
 $7$ & &$6 \times 10^{-5}$ & $1263.25$ & UK & Cyprus & & $4 \times 10^{-33}$ & $82175.1$ & China & Taiwan 
 \\ 
 $8$ & &$6 \times 10^{-5}$ & $727.11$ & Uganda & Kenya & & $8 \times 10^{-30}$ & $3893.1$ & USA & Honduras 
 \\ 
 $9$ & &$6 \times 10^{-5}$ & $1580.3$ & USA & Nicaragua & & $3 \times 10^{-28}$ & $62399.9$ & Austria & Germany 
 \\ 
 $10$ & &$8 \times 10^{-5}$ & $7572$ & USA & Ecuador & & $8 \times 10^{-26}$ & $315362$ & USA & China 
 \\ 
 $11$ & &$9 \times 10^{-5}$ & $1642.65$ & Benin & China & & $9 \times 10^{-26}$ & $103930$ & China & South Korea 
 \\ 
 $12$ & &$9 \times 10^{-5}$ & $3326.3$ & USA & Guatemala & & $9 \times 10^{-26}$ & $19399.3$ & USA & Israel 
 \\ 
 $13$ & &$1 \times 10^{-3}$ & $4062.74$ & Honduras & USA & & $9 \times 10^{-25}$ & $7572$ & USA & Ecuador 
 \\ 
 $14$ & &$1 \times 10^{-3}$ & $508.5$ & USA & Haiti & & $5 \times 10^{-24}$ & $143421$ & Mexico & USA 
 \\ 
 $15$ & &$2 \times 10^{-3}$ & $1458.28$ & Zambia & South Africa & & $1 \times 10^{-23}$ & $38642.4$ & Germany & Austria 
 \\ 
 $16$ & &$2 \times 10^{-3}$ & $187.33$ & St. Kitts \& Nevis & USA & & $1 \times 10^{-22}$ & $106105$ & Germany & Netherlands 
 \\ 
 $17$ & &$2 \times 10^{-3}$ & $2282.3$ & USA & Sri Lanka & & $5 \times 10^{-22}$ & $27804.6$ & Germany & Czech Rep 
 \\ 
 $18$ & &$2 \times 10^{-3}$ & $1056.25$ & Mozambique & South Africa & & $1 \times 10^{-21}$ & $8822.4$ & USA & Trinidad \& Tobago 
 \\ 
 $19$ & &$2 \times 10^{-3}$ & $459.05$ & India & Nepal & & $4 \times 10^{-21}$ & $27330.9$ & Ireland & UK 
 \\ 
 $20$ & &$2 \times 10^{-3}$ & $193.22$ & China & Burkina Faso & & $5 \times 10^{-21}$ & $19308.9$ & Portugal & Spain 
 \\ 
\cline{3-6} \cline{8-11}

\end{tabular}
\caption{\label{tab:1} List of the top $20$ most 
relevant connections of the World
  Trade Web according to the GloSS (left) and the Disparity
  filter (right), respectively. The weights are evaluated in millions
  of dollars. The edges selected by the Disparity filter carry on
  average much larger weights and have far lower p-values than those picked by GloSS. }
\end{table*}

\subsection{Tests on real networks}
\label{secreal}

Here we show some applications of our filtering procedure to real
weighted networks. First we focus our attention on the most
``significant'' weights of the network. For this purpose we take the
World Trade Web (WTW)~\cite{serrano03}, i.e. the network of trade relationships of
world countries. Vertices represent the countries and edges are
directed and weighted by the money flow running from any two countries
to the other (import/export). The WTW is very useful to study
propagation of economic crises and has been thoroughly investigated
in the last years~\cite{serrano03,garlaschelli04,garlaschelli05}. Data are freely available~\cite{barbieri08,barbieri09}.
The data we considered refer to the year $2006$: the network has $189$
vertices and $12\,705$ edges.
In Fig.~\ref{fig:trade} we show the $50$ most significant edges, selected with the GloSS (left) and the Disparity
(right) filter, respectively. We see that the results are quite different,
even if some of the edges coincide. In particular, the GloSS filter
is more likely to capture connections involving smaller/poorer
countries than the Disparity filter, which selects more frequently 
larger countries and trade exchanges. This is manifest in Table~\ref{tab:1},
where we list the top $20$ edges, along with their weights and p-values.

\begin{figure*}
\begin{center}
\includegraphics[width=.98\textwidth]{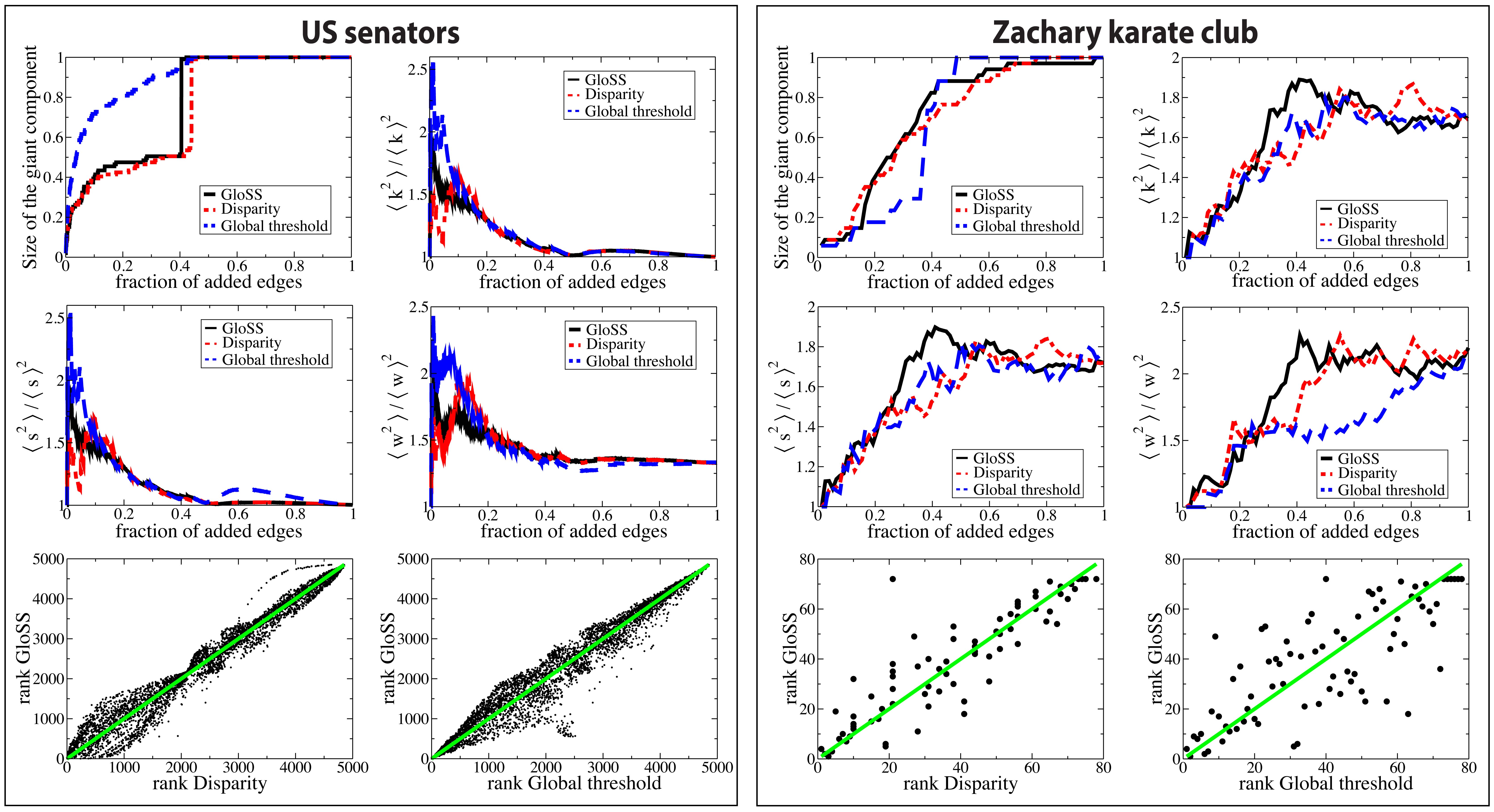}
\caption{\label{fig:undirected} (Color online) Applications of filtering techniques
  on two real weighted undirected networks: a network of US senators (left)
  and Zachary's karate club (right). For each network we show the size of the
  largest connected component and the heterogeneity parameter of the
  degree ($k$), strength ($s$) and edge weight ($w$) distributions as a function of the number of edges
  added to the system (in decreasing order of relevance). The
  continuous line stands for the results of the GloSS filter, the
  dot-dashed line for those of the Disparity filter, the dashed line
  for global thresholding. In addition,
we show scatter plots of the edge rankings estimated by the
GloSS filtering technique and the other two
considered here: Disparity and
global thresholding.}
\end{center}
\end{figure*}

Interesting economic relations, revealed as anomalous by the GloSS filter,
are those between China and North Korea and also those
relating China to Togo, Burkina Faso and Benin.
While the existence of an anomalous connection
between China and North Korea can be explained in terms of simple
political reasons, the relations of China
with the African countries have deeper economic foundations
based on agreements on trade, economic and technological cooperation.
Particularly relevant economic relations are also those established
between Australia and Papua New Guinea, 
between Italy and Albania and between France and Gabon. 
Papua New Guinea became independent from 
Australia only in $1975$, but
its economic development is still controlled by Australia.
After the collapse of communism in Albania ($1991$), 
a mass exodus of refugees moved
to Italy. Albanians form nowdays 
one of the largest foreign
communities in Italy and strong trade relationships are present
between the two countries. Gabon was a colony of France
up to $1960$, but still maintain exclusive  
political and economic relationships with France.

We now proceed with a more systematic study of the importance of the
selected weights for the structure of the network. Since the goal is
to reduce the information of the system by keeping as many as possible
of its features, one may wonder how many edges, picked in descending
order of significance, are necessary to
reproduce the most important features of the original weighted graph. 
For instance, how many edges are needed to form a connected graph?
This test has been suggested in Ref.~\cite{serrano09}. In addition, we
wish to check when the distributions of the vertex degrees, vertex
strengths and edge weights are restored. Since it is hard to verify
the match of two distributions, while it is far easier to compare two numbers, 
we limit the comparison to an important property of a distribution,
the {\it heterogeneity parameter}, expressing the dispersion of
the distribution around its average. For a variable $x$ with a 
certain probability
distribution, the heterogeneity parameter is defined as the ratio of the
second moment of the distribution by the square of the first moment: $\langle
x^2\rangle/\langle x\rangle^2$. Our tests consist then in adding edges 
until the heterogeneity parameters of the distributions of the reduced
network reach those of the original network and remain stable until
the last edges are added. In Appendix~\ref{appkl} we use an
alternative measure for the comparison of distributions: the
Kullback-Leibler divergence~\cite{kullback51}. We carried out the tests by using three
different filtering techniques: GloSS, Disparity and global thresholding. 
We also compare the rankings produced by our filtering method with
those obtained with the other techniques 
to estimate their correlation.

We start with two undirected graphs: a network of US senators~\cite{poole07} and
the karate club network of Zachary~\cite{zachary}. The first is a
network with $99$ vertices, corresponding to members of the 109th Senate of
the United States that served for the full two-years term. The weight of
the edge between a pair of senators is
weighted by the number of times they have voted
in the same way (the total number of edges is $4\,851$). The data are
freely available from {\tt http://voteview.com}. 
\begin{figure*}
\begin{center}
\includegraphics[width=.98\textwidth]{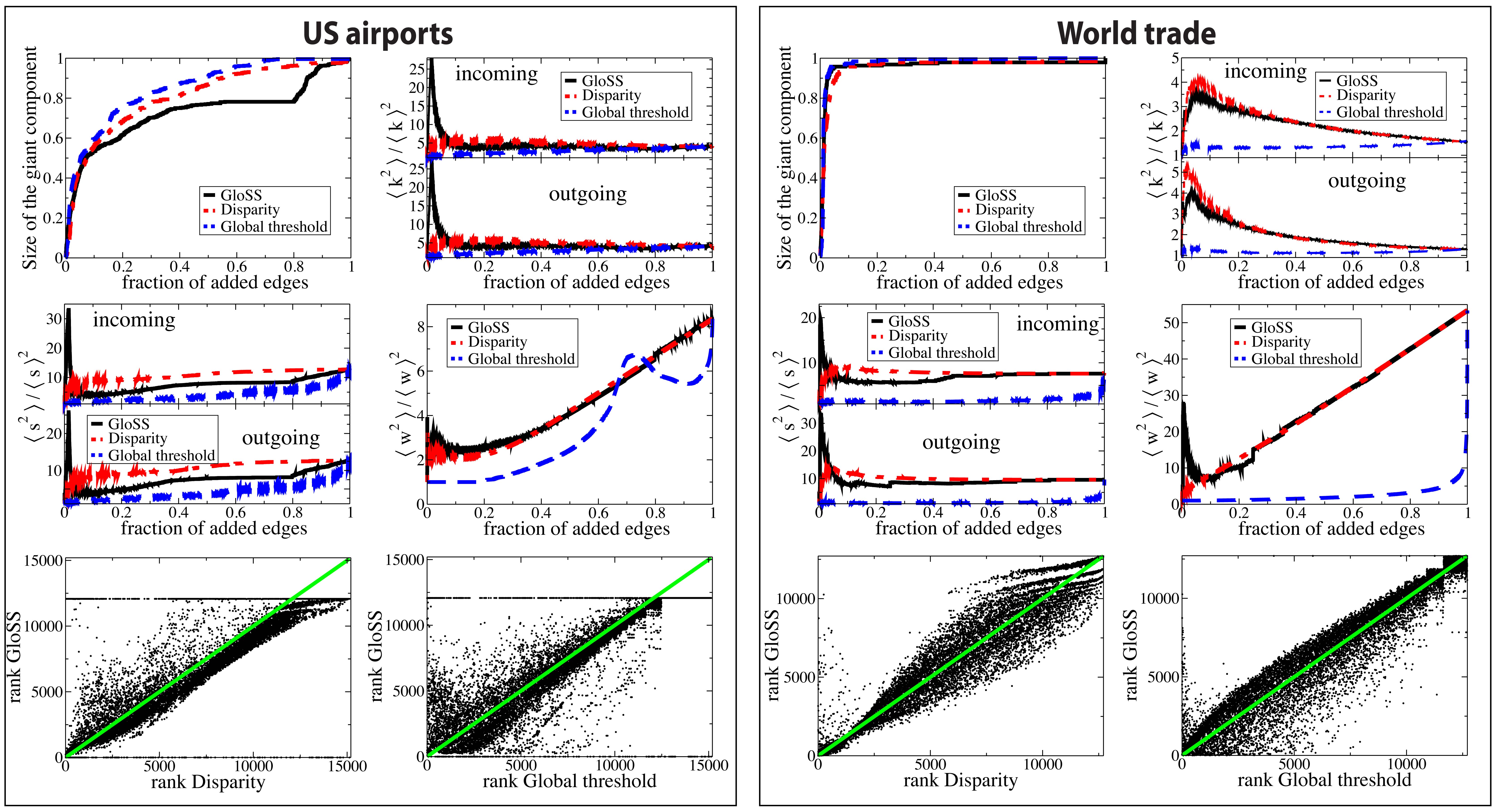}
\caption{\label{fig:directed} (Color online) Applications of filtering techniques
  on two real weighted directed networks: the US airport network
  (left) and
  the World Trade Web (right). The panels are
  analogous as those of Fig.~\ref{fig:undirected}, although those
  relative to the degree and strength distributions 
  are split to account for the two possible edge
  directions (incoming and outgoing). The
  continuous line stands for the results of the GloSS filter, the
  dot-dashed line for those of the Disparity filter, the dashed line
  for global thresholding.}
\end{center}
\end{figure*}
Naturally,
senators of the same party (Republican or Democratic) are more likely
to vote together than senators of different parties. Consequently, the
distribution of edge weights is bimodal, with two groups of values
corresponding to edges joining Republican or Democratic senators and
to edges joining Republicans to Democrats. Zachary's karate club
network consists of $34$ vertices and $78$ edges, corresponding to the
members of a karate club in the USA and their social relationships. It
has become quite popular lately as it is frequently used as
benchmark to test algorithms for community
detection~\cite{fortunato2010}. In Fig.~\ref{fig:undirected} we show
the results of our analysis of both graphs. The performances of the
GloSS and Disparity filters are rather similar. For the
senators network we see that after adding about $40\%$ of the edges
the reduced network acquires the features of the original one. In this
case, there is a strong correlation between the GloSS filter and
global thresholding when it comes to selecting the most relevant
edges.
This is due to the fact that 
the senator network is almost fully connected and its weight distribution is
bimodal (as opposed to the typically broad distributions observed in
many systems). Under the null model assumption of random assignments of
weights (from the given bimodal distribution), the larger weights
between members of the same party are
more likely to be deemed relevant by the GloSS filter.

\begin{figure*}[htb]
\begin{center}
\includegraphics[width=.98\textwidth]{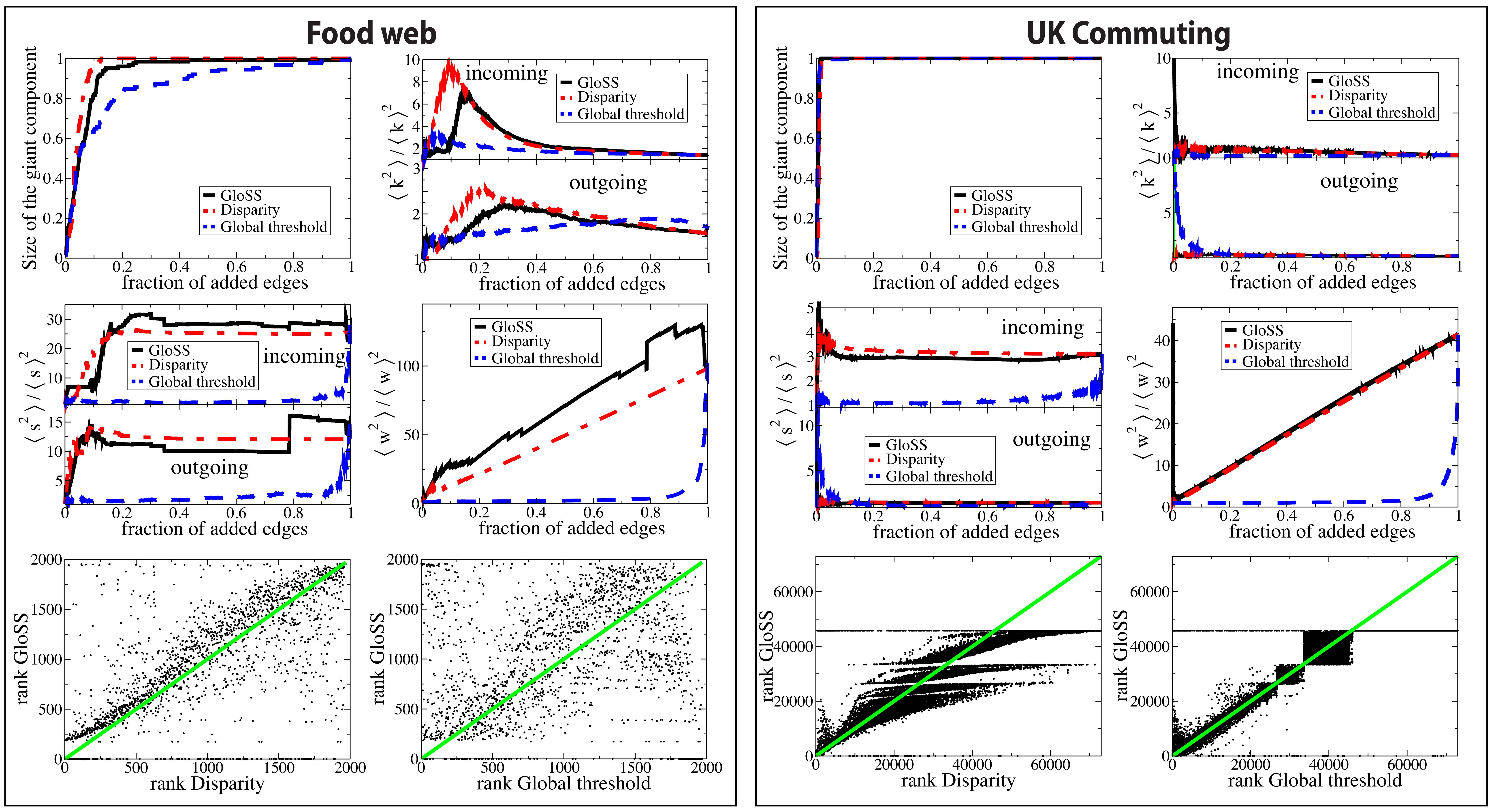}
\caption{\label{fig:S1} (Color online) Applications of filtering techniques
  on two real weighted directed networks:
  the food web of Florida Bay in the dry season and
  the commuting network between cities in UK. The panels
  report the same analyses as those of Fig.~\ref{fig:directed}. The
  continuous line stands for the results of the GloSS filter, the
  dot-dashed line for those of the Disparity filter, the dashed line
  for global thresholding.}
\end{center}
\end{figure*}

Finally, we discuss applications to directed networks. We take 
four datasets: the World Trade Web (WTW), the air transportation network of the USA, the Florida Bay ecosystem in the dry
season~\cite{ulanowicz98} and a network of commuting in the UK. 
The WTW has been described at the beginning of this subsection.
Data on the US air transportation network 
can be downloaded from the
Bureau of Transportation Statistics (US government) ({\tt http://www.bts.gov}).
Vertices are US airports and edges are weighted by the number of
passengers transported along the corresponding routes in the year 2000. Our
network has $664$ vertices and $15\,132$ edges.
The food web of Florida Bay entails the trophic interactions 
between species, weighted by carbon transfers 
from one species to another. The network has been constructed within the
ATLSS Project of the University of Maryland ({\tt http://www.cbl.umces.edu/atlss.html})
The species are $125$,  their interactions $1\,969$.
The network of commuting is composed of $376$ vertices,
representing local authorities, geographical divisions covering the 
territories of England and Wales. 
Each of the $72\,954$ directed edges corresponds to a flow of
commuters between the local authority of 
origin and that of destination with a weight accounting for the number
of commuters per day. 
The data come from the $2001$ UK census, where the local authority of
residence and of work/study 
is registered for a significative part of the British population. The
database can be accessed 
online at the site of the Office for National Statistics
{\tt http://www.ons.gov.uk/census}.

In Fig.~\ref{fig:directed} we show the results of our analysis for the
WTW and the US airport network,
following the same scheme as in Fig.~\ref{fig:undirected}. The results
for the food web and the network of commuting are reported in Fig.~\ref{fig:S1}.
We remark again a substantial similarity between the GloSS and the
Disparity filter. This seems to be odd, as the two
filtering procedures are very different in their selection of
the most significant edges, as we have shown in Fig.~\ref{fig:trade}
and Table~\ref{tab:1}. What emerges from
Figs.~\ref{fig:undirected},~\ref{fig:directed} and~\ref{fig:S1} is
that if a sizable fraction of edges are picked,
both filters select mostly the same weights, so after a while the
reduced descriptions of the network would match or become very similar.
On the other hand global thresholding is clearly inadequate to
catch the main properties of the original network, for it requires many
more edges to recover them, as already pointed out in
Ref.~\cite{serrano09}.

\begin{figure}[htb]
\begin{center}
\includegraphics[width=\columnwidth]{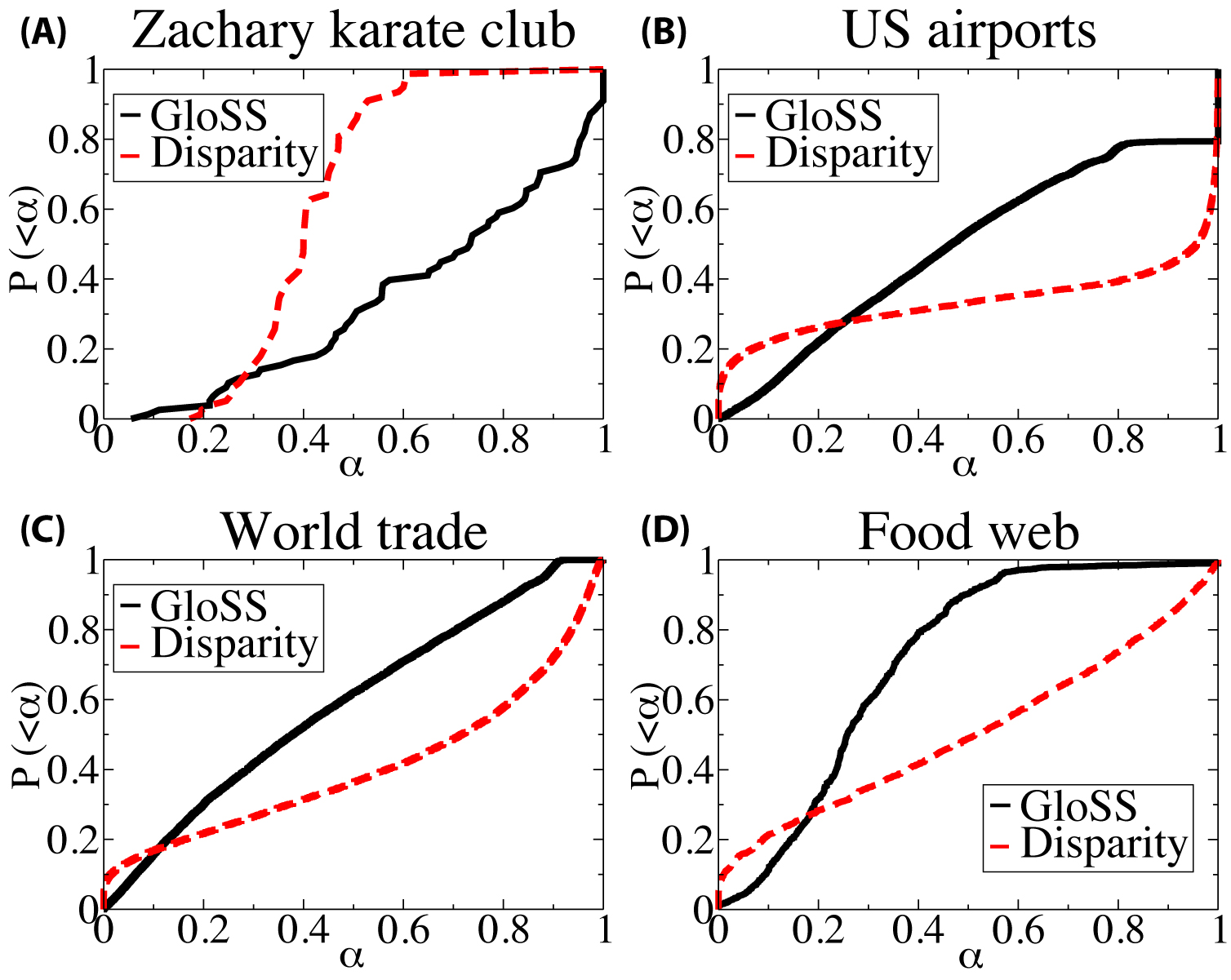}
\caption{\label{fig6} (Color online) Cumulative distribution $P\left(<\alpha\right)$
of the significance level $\alpha$
for weights taken from the observed distribution of some
of the real networks we considered. The continuous line corresponds to
the GloSS filter, the dashed line to the Disparity filter.}
\end{center}
\end{figure}

We close the section by performing a study analogous to that reported
in Fig.~\ref{fig:null_model}, but for some of the real networks
examined here (Fig.~\ref{fig6}). For the GloSS filter (continuous lines) 
we find different patterns than that expected for the null model,
in which all p-values are equally probable. Only for the US airports
the p-values have roughly the same probability, up until $\alpha\sim
0.8$. For Zachary's karate club, the WTW and the food web, there are
significant differences with respect to the null model. The
Disparity filter (dashed lines) displays a markedly different
behavior: with the exception of Zachary's karate club, very low
$\alpha$ values are much more frequent than found by the GloSS filter.

\section{Conclusions}

Filtering the information of complex weighted networks is crucial both
to detect the most relevant connections and to be able to process a system that
is often too large for many analytical tools to work efficiently. In
this paper we have presented the first filtering technique based on a
consistent global null model, preserving both the distribution of edge
weights and the full topology of the graph. The recipe is by no means
unique and it would not be difficult to propose alternatives with
slight modifications of the main ingredients. In fact, filters are as
arbitrary as the notion of ``relevant information'' is, so objective
comparisons of different strategies are unfeasible. 
Still, there are situations in which the answer of the filter is
intuitive. For instance, if weights are
independently and identically distributed among the edges, there
should be no anomalous fluctuations and, consequently, the $p$-values
of the edges should be homogeneously distributed. 
We have seen that our GloSS
filter indeed quantifies
the correct statistical significance in such instances, while other
techniques have problems.

Tests on real weighted networks show that the GloSS filter is capable
of subsuming the basic information about the system in a
fairly small fraction of the edges, especially 
the multiscale structure of both the topology and the edge weights. 
While we have put some emphasis on networks with heterogeneous distributions of features,
we remark that our procedure is very general and it applies as well to cases in which
distributions are peaked, as we have seen for the network
of US senators. 
The significance of the edges is not so strongly
correlated with their weights like for other techniques, so we are
able to obtain potentially relevant information also from the
vertices with low strength and degree and, consequently, a more balanced tradeoff
between topology and weights.

Therefore we believe that the GloSS filter is a valuable tool for the
analysis of networked datasets. The procedure is implemented in a freely
downloadable software ({\tt http://filrad.homelinux.org/resources}).

\appendix
\section{Numerical implementation of the GloSS filter}
The evaluation of the Fourier transform (and antitransform)
can be performed by using a Fast Fourier Transform (FFT) algorithm.
This requires as input a binned version of the weight
distribution, where the number of bins $b$ must be a power of $2$. 
The range of values
we are interested in is $\left[0,S\right]$, where
$S=k_{max}w_{max}$ is the product of the maximal
degree and the maximal weight observed
in the network. A proper number of bins
is needed in order to be able to distinguish
different weight values: if $\delta w$ is the
minimum value of the difference among
all pairs of unequal weights in the network, we set
 $Q = \lceil \log_2 \left(S/\delta w\right) \rceil$
and perform the linear binning of  $P_{obs}\left(w\right)$
over $b=2^Q$ bins. We implement
our filtering technique by calculating the Fourier transform
of the weight distribution and
all its powers up to $k_{max}$. For each resulting expression
we obtain the Fourier antitransform and finally
compute the $p$-values of all edges according
to Eq.~(\ref{eq:pval}). The complexity of the various
stages of our algorithm can be simply estimated: $b \, \log_2\left(b\right)$ is
the typical complexity for calculating
the Fourier transform or antitransform; computing
the powers of the Fourier transform requires a time
which grows as $b \, k_{max}$; deriving
the inverse of the Fourier transform for
each power scales as $k_{max} \, b\, \log_2\left(b\right)$;
evaluating the statistical significance for each of the $M$ edges
in the network goes as $M \, b$. Since in general $M \gg k_{max}$,
the computational complexity of the whole filtering technique proposed
in this paper is $M \, b = M \, 2^Q$.

\section{Matching the backbone and the original graph}
\label{appkl}

\begin{figure}[htb]
\begin{center}
\includegraphics[width=\columnwidth]{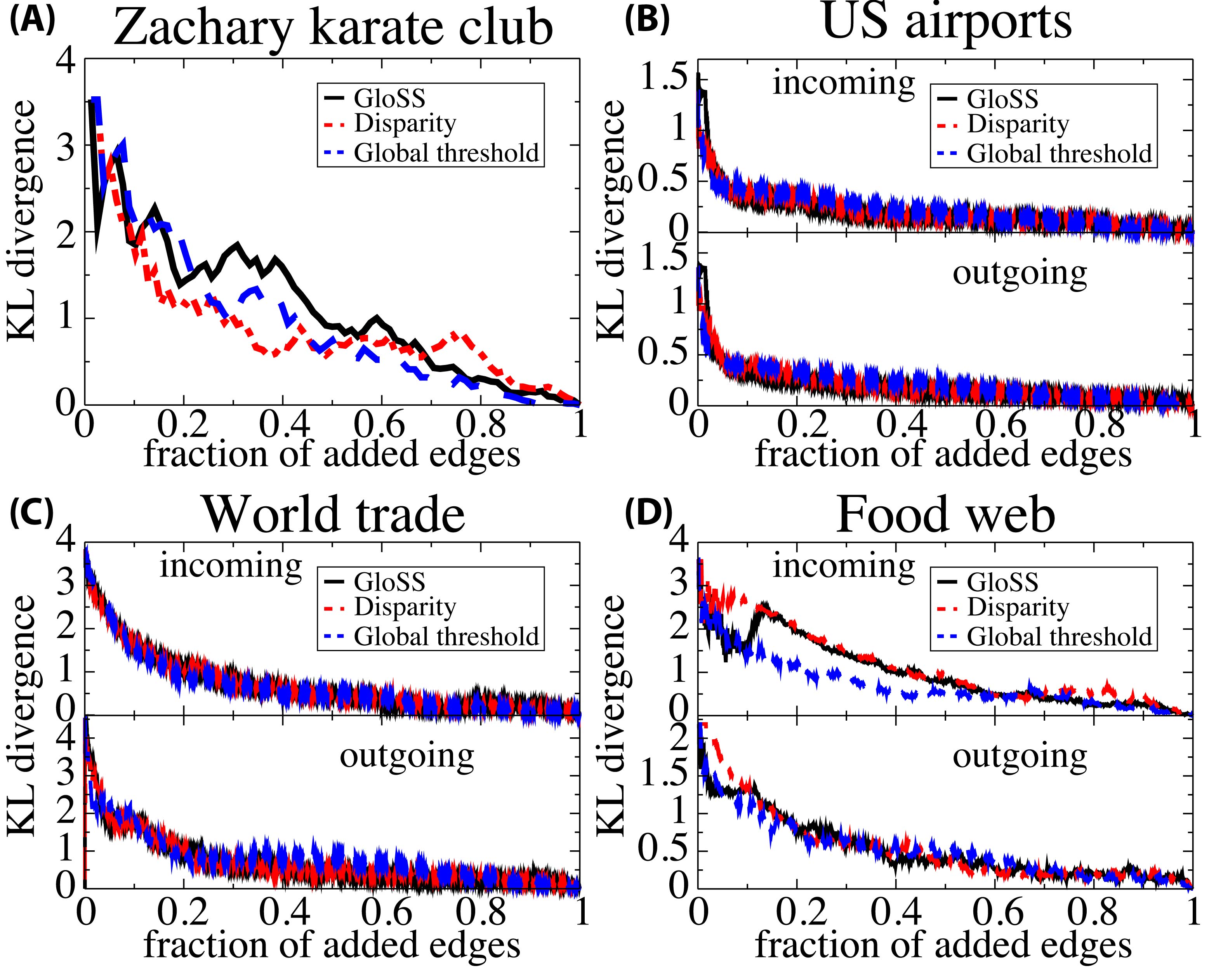}
\caption{\label{fig7} (Color online) Each panel shows the difference
  between the degree distributions of the filtered networks and those
  of the original systems, measured by the Kullback-Leibler divergence, for the GloSS
  filter (continuous line), the Disparity filter (dot-dashed line) and
  global thresholding (dashed line). For directed
  graphs the panels are split in two halves, referring to the
  in-degree and the out-degree distributions, respectively.}
\end{center}
\end{figure}

In Section~\ref{secreal} we have compared the distribution of local
properties of the backbone with that of the original graph, to check
how many edges are needed to reproduce the basic features of the graph
at study. For this purpose we have compared the heterogeneity
parameters of corresponding distributions as a function of the fraction of
added edges. To give more robustness to our results, we consider here
an alternative measure for the comparison of distributions, the
Kullback-Leibler (KL) divergence~\cite{kullback51}, a well-known
measure in information theory. The results are
shown in Fig.~\ref{fig7} for four real networks. As we had found in Section~\ref{secreal},
there is little difference between the GloSS and the Disparity
filters, while global thresholding follows slightly different trends.

\begin{acknowledgments}
We thank A. Lancichinetti, M. \'A. Serrano and A. Vespignani for stimulating
discussions. S. F. and J.J.R. gratefully acknowledge funding from
ICTeCollective and Dynanets, respectively.  These are projects
of the Future and Emerging Technologies (FET) programme within the
Seventh Framework 
Programme for Research of the European Commission, under FET-Open
grant numbers 238597 (ICTeCollective) and 233847 (Dynanets).

\end{acknowledgments}

\end{document}